\documentclass[twocolumn]{aastex63}
\pdfoutput=1 
\usepackage{appendix}
\usepackage{graphicx}
\usepackage{amsmath,amstext}
\usepackage[T1]{fontenc}
\usepackage[figure,figure*]{hypcap}

\newcommand{\kms}{\,km\,s$^{-1}$}

\newcommand{\G}{\mathcal{G}}

\newcommand{\appropto}{\mathrel{\vcenter{\offinterlineskip\halign{\hfil$##$\cr\propto\cr\noalign{\kern2pt}\sim\cr\noalign{\kern-2pt}}}}}

\newcommand{\feh}{\ensuremath{\left[{\rm Fe}/{\rm H}\right]}}
\newcommand{\teff}{\ensuremath{T_{\rm eff}}}
\newcommand{\logg}{\ensuremath{\,{\rm log}\,{g}}}

\newcommand{\msun}{\ensuremath{\,{\rm M_\Sun}}}
\newcommand{\rsun}{\ensuremath{\,{\rm R_\Sun}}}

\newcommand{\mj}{\ensuremath{\,{\rm M_{\rm J}}}}
\newcommand{\rj}{\ensuremath{\,{\rm R_{\rm J}}}}
\newcommand{\degree}{\ensuremath{\,^{\circ}}}

\newcommand{\fdeg}{\ensuremath{\,^{\!\!\circ}}}
\newcommand{\PSUAA}{Department of Astronomy \& Astrophysics, 525 Davey Laboratory, The Pennsylvania State University, University Park, PA, 16802, USA}
\newcommand{\PSUCEHW}{Center for Exoplanets and Habitable Worlds, 525 Davey Laboratory, The Pennsylvania State University, University Park, PA, 16802, USA}
\newcommand{\PSETI}{Penn State Extraterrestrial Intelligence Center, 525 Davey Laboratory, The Pennsylvania State University, University Park, PA, 16802, USA}

\graphicspath{{./}{figures/}}

\received{2021 October 17}
\revised{2022 January 18}
\accepted{2022 January 18}
\submitjournal{ApJL}
\shorttitle{Low Obliquity for WASP-148}
\shortauthors{Wang et al.}

\begin{document}

\title{The Aligned Orbit of WASP-148b, the Only Known Hot Jupiter with a Nearby Warm Jupiter Companion, from NEID and HIRES}

\author[0000-0002-0376-6365]{Xian-Yu Wang} 
\affil{National Astronomical Observatories, Chinese Academy of Sciences, Beijing 100012, China}
\affil{University of Chinese Academy of Sciences, Beijing, 100049, China}

\author[0000-0002-7670-670X]{Malena Rice} 
\affil{Department of Astronomy, Yale University, New Haven, CT 06511, USA}
\affil{NSF Graduate Research Fellow}

\author[0000-0002-7846-6981]{Songhu Wang}
\affil{Department of Astronomy, Indiana University, Bloomington, IN 47405}

\correspondingauthor{Songhu Wang}
\email{sw121@iu.edu}

\author[0000-0002-5668-243X]{Bonan Pu} 
\affil{Department of Astronomy, Center for Astrophysics and Planetary Science, Cornell University, Ithaca, NY 14850, USA}

\author[0000-0001-7409-5688]{Guðmundur Stefánsson} 
\affil{Henry Norris Russell Fellow}
\affil{Princeton University, Department of Astrophysical Sciences, 4 Ivy Lane, Princeton, NJ 08540, USA}

\author[0000-0001-9596-7983]{Suvrath Mahadevan} 

\affil{\PSUAA}
\affil{\PSUCEHW}

\author[0000-0002-0015-382X]{Brandon Radzom}
\affil{Department of Astronomy, Indiana University, Bloomington, IN 47405}

\author[0000-0002-8965-3969]{Steven Giacalone} 
\affil{Department of Astronomy, University of California Berkeley, Berkeley, CA 94720, USA}
\author[0000-0001-8037-1984]{Zhen-Yu Wu} 
\affil{National Astronomical Observatories, Chinese Academy of Sciences, Beijing 100012, China}
\affil{University of Chinese Academy of Sciences, Beijing, 100049, China}
\author[0000-0002-0792-3719]{Thomas M. Esposito}
\affil{SETI Institute, Carl Sagan Center, 189 Bernardo Avenue, Mountain View, CA 94043, USA}
\affil{Department of Astronomy, University of California Berkeley, Berkeley, CA 94720, USA}
\affil{Unistellar SAS, 19 Rue Vacon, 13001 Marseille, France}

\author[0000-0002-4297-5506]{Paul A.\ Dalba}  
\affil{NSF Astronomy and Astrophysics Postdoctoral Fellow}

\affil{Department of Astronomy and Astrophysics, University of California, Santa Cruz, CA 95064, USA}
\affil{Department of Earth and Planetary Sciences, University of California Riverside, Riverside, CA 92521, USA}

\author[0000-0001-7801-7425]{Arin Avsar} 
\affil{Unistellar SAS, 19 Rue Vacon, 13001 Marseille, France}
\affil{Department of Astronomy, University of California Berkeley, Berkeley, CA 94720, USA}

\author[0000-0002-6153-3076]{Bradford Holden} 
\affil{UCO/Lick Observatory, Department of Astronomy and Astrophysics, University of California at Santa Cruz,Santa Cruz, CA 95064}

\author{Brian Skiff}
\affil{Lowell Observatory, 1400 West Mars Hill Road, Flagstaff, AZ 86001, USA}

\author[0000-0003-0628-7017]{Tom Polakis} 
\affil{Lowell Observatory, 1400 West Mars Hill Road, Flagstaff, AZ 86001, USA}

\author{Kevin Voeller}
\affil{Science Department, Garden Grove High School, 11270 Stanford Avenue, Garden Grove, CA 92840}

\author[0000-0002-9632-9382]{Sarah E. Logsdon}
\affil{NSF's National Optical-Infrared Astronomy Research Laboratory, 950 N. Cherry Ave., Tucson, AZ 85719, USA}

\author[0000-0003-3906-9518]{Jessica Klusmeyer} 
\affil{NSF's National Optical-Infrared Astronomy Research Laboratory, 950 N. Cherry Ave., Tucson, AZ 85719, USA}

\author[0000-0001-9580-4869]{Heidi Schweiker} 
\affil{NSF's National Optical-Infrared Astronomy Research Laboratory, 950 N. Cherry Ave., Tucson, AZ 85719, USA}

\author[0000-0001-9424-3721]{Dong-Hong Wu}
\affil{Department of Physics, Anhui Normal University, Wuhu Anhui, 241000, PR, China}

\author[0000-0001-7708-2364]{Corey Beard}
\affil{Department of Physics and Astronomy, University of California Irvine, Irvine, CA 92697, USA}

\author[0000-0002-8958-0683]{Fei Dai}
\affil{Division of Geological and Planetary Sciences 1200 E California Blvd, Pasadena, CA 91125}

\author[0000-0001-8342-7736]{Jack Lubin} 
\affil{Department of Physics and Astronomy, University of California Irvine, Irvine, CA 92697, USA}

\author[0000-0002-3725-3058]{Lauren M. Weiss}
\affil{Department of Physics, University of Notre Dame, Notre Dame, IN 46556, USA}

\author[0000-0003-4384-7220]{Chad F. Bender}
\affil{Steward Observatory, The University of Arizona, 933 N. Cherry Avenue, Tucson, AZ 85721, USA}

\author[0000-0002-6096-1749]{Cullen H. Blake} 
\affil{Department of Physics and Astronomy, University of Pennsylvania, 209 S. 33rd Street, Philadelphia, PA 19104, USA}

\author[0000-0001-8189-0233]{Courtney D. Dressing} 
\affil{Department of Astronomy, University of California Berkeley, Berkeley, CA 94720, USA}

\author[0000-0003-1312-9391]{Samuel Halverson} 
\affil{Jet Propulsion Laboratory, California Institute of Technology, 4800 Oak Grove Drive, Pasadena, California 91109} 

\author[0000-0002-1664-3102]{Fred Hearty}

\affil{\PSUAA}
\affil{\PSUCEHW}

\author[0000-0001-8638-0320]{Andrew W. Howard} 
\affil{Department of Astronomy, California Institute of Technology, Pasadena, CA 91125, USA}

\author[0000-0001-8832-4488]{Daniel Huber} 
\affil{Institute for Astronomy, University of Hawai`i, 2680 Woodlawn Drive, Honolulu, HI 96822, USA}

\author[0000-0002-0531-1073]{Howard Isaacson} 
\affil{Department of Astronomy, University of California Berkeley, Berkeley, CA 94720, USA}
\affil{Centre for Astrophysics, University of Southern Queensland, Toowoomba, QLD, Australia}

\author[0000-0003-0711-7992]{James A. G. Jackman} 
\affil{School of Earth and Space Exploration, Arizona State University, Tempe, AZ 85287, USA}

\author[0000-0003-4450-0368]{Joe Llama} 
\affil{Lowell Observatory, 1400 West Mars Hill Road, Flagstaff, AZ 86001, USA}

\author[0000-0003-0241-8956]{Michael W. McElwain} 
\affil{Exoplanets and Stellar Astrophysics Laboratory, NASA Goddard Space Flight Center, Greenbelt, MD 20771, USA}

\author[0000-0002-2488-7123]{Jayadev Rajagopal} 
\affil{NSF's National Optical-Infrared Astronomy Research Laboratory, 950 N. Cherry Ave., Tucson, AZ 85719, USA}

\author[0000-0001-8127-5775]{Arpita Roy} 
\affil{Space Telescope Science Institute, 3700 San Martin Drive, Baltimore, MD 21218, USA}
\affil{Department of Physics and Astronomy, Johns Hopkins University, 3400 N Charles St, Baltimore, MD 21218, USA}

\author[0000-0003-0149-9678]{Paul Robertson} 
\affil{Department of Physics and Astronomy, University of California Irvine, Irvine, CA 92697, USA}

 \author[0000-0002-4046-987X]{Christian Schwab} 
 \affil{Centre for Astronomy, Astrophysics and Astrophotonics, Macquarie University – Sydney, NSW 2109, Australia}
 
\author[0000-0002-7260-5821]{Evgenya L. Shkolnik}
\affil{School of Earth and Space Exploration, Arizona State University, Tempe, AZ 85287, USA}

\author[0000-0001-6160-5888]{Jason T.\ Wright}
\affil{\PSUAA}
\affil{\PSUCEHW}
\affil{\PSETI}

\author[0000-0002-3253-2621]{Gregory Laughlin} 
\affil{Department of Astronomy, Yale University, New Haven, CT 06511, USA}



\begin{abstract}
\noindent

We present spectroscopic measurements of the Rossiter-McLaughlin effect for WASP-148b, the only known hot Jupiter with a nearby warm-Jupiter companion, from the WIYN/NEID and Keck/HIRES instruments. This is one of the first scientific results reported from the newly commissioned NEID spectrograph, as well as the second obliquity constraint for a hot Jupiter system with a close-in companion, after WASP-47. WASP-148b is consistent with being in alignment with the sky-projected spin axis of the host star, with $\lambda=-8\degree.2^{{+8\fdeg.7}}_{-9\fdeg.7}$. The low obliquity observed in the WASP-148 system is consistent with the orderly-alignment configuration of most compact multi-planet systems around cool stars with obliquity constraints, including our solar system, and may point to an early history for these well-organized systems in which migration and accretion occurred in isolation, with relatively little disturbance. By contrast, previous results have indicated that high-mass and hot stars appear to more commonly host a wide range of misaligned planets: not only single hot Jupiters, but also compact systems with multiple super-Earths. We suggest that, to account for the high rate of spin-orbit misalignments in both compact multi-planet and isolated-hot-Jupiter systems orbiting high-mass and hot stars, spin-orbit misalignments may be caused by distant giant planet perturbers, which are most common around these stellar types.

\end{abstract}



\section{Introduction} \label{sec:intro}

While hot Jupiters are the most observationally accessible population of exoplanets, their origins remain unclear. Two primary dynamical patterns in the hot Jupiter population have emerged from over two decades of observational efforts. First, most hot Jupiters are not accompanied by nearby planet companions \citep{Steffen2012, Huang2015, WangX2021}. Second,  a significant fraction of hot-Jupiter orbits are misaligned with their host stars' spin axes (as reviewed by \citealt{Winn2015}).

High-eccentricity migration, which strips a hot Jupiter of primordial neighboring planets and leaves the system misaligned, naturally reproduces both of these patterns. As a result, it has generally been viewed as the primary mechanism that delivers hot Jupiters to their current locations (as reviewed by \citealt{Dawson2018}).

If true, this hypothesis would suggest that hot Jupiters' loneliness should be associated with their host stars' obliquities: significantly non-zero spin-orbit angles should be confined to isolated hot Jupiters. The existence of a hot Jupiter and a nearby companion planet in the same system essentially precludes a dynamically violent history. As a result, hot-Jupiter systems with one or more nearby planets should have low stellar obliquities indicative of a dynamically quiescent formation route.

In this light, the WASP-148 system has a special importance as one of only four known systems containing a hot Jupiter that is also part of a compact multi-planet system (the other three are WASP-47, \citealt{Becker2015}; Kepler-730, \citealt{Canas2019}; and TOI-1130, \citealt{Huang2020}). WASP-148 is a G star with $\teff = 5,437\pm21$ K and $M_*=0.97^{+0.056}_{-0.057} \ \rm M_{\odot}$, hosting two confirmed giant planets at $P= 8.803544\pm0.000021$ days and $P=34.527\pm0.024$ days \citep{hebrard2020discovery}. The inner planet WASP-148b with $M_{\rm b} = 0.354^{+0.055}_{-0.050} \mj$, transits its host star, while the outer companion with $M_{\rm c}{\rm sin}i = 0.408_{-0.087}^{+0.127} \mj$ does not transit but is suspected to be coplanar based on the global analysis of transit timing variations and radial-velocity (RV) data \citep{Maciejewski2020}.

We present the small sky-projected spin-orbit angle ($\lambda = -8\degree.2_{-9\fdeg.7}^{+8\fdeg.7}$) of WASP-148 observed through the Rossiter-McLaughlin effect \citep[R-M:][]{rossiter1924detection, mclaughlin1924some} across two separate transits of WASP-148b, as the second result in our stellar obliquities survey \citep{Rice2021} and the second obliquity constraint for a hot Jupiter system with a close-in companion, after WASP-47 \citep{Sanchis2015}. Our two R-M measurements were obtained using the NEID spectrograph \citep{schwab2016design} on the WIYN 3.5-meter telescope and the HIRES spectrograph \citep{vogt1994hires} on the 10-meter Keck I telescope. The derived obliquity provides a rare opportunity to probe the correlation between hot Jupiters' loneliness and their host stars' obliquities, as well one of the first demonstrations of the precise radial-velocity data collected with NEID.

In what follows, we describe our photometric and RV observations (\S 2), the parameterized model used to determine the stellar obliquity (\S 3), and the possible implications of our results (\S 4). 
\vspace{0.5cm}
\section{Observations}

\begin{figure*}
   \includegraphics[width = 1.0\textwidth]{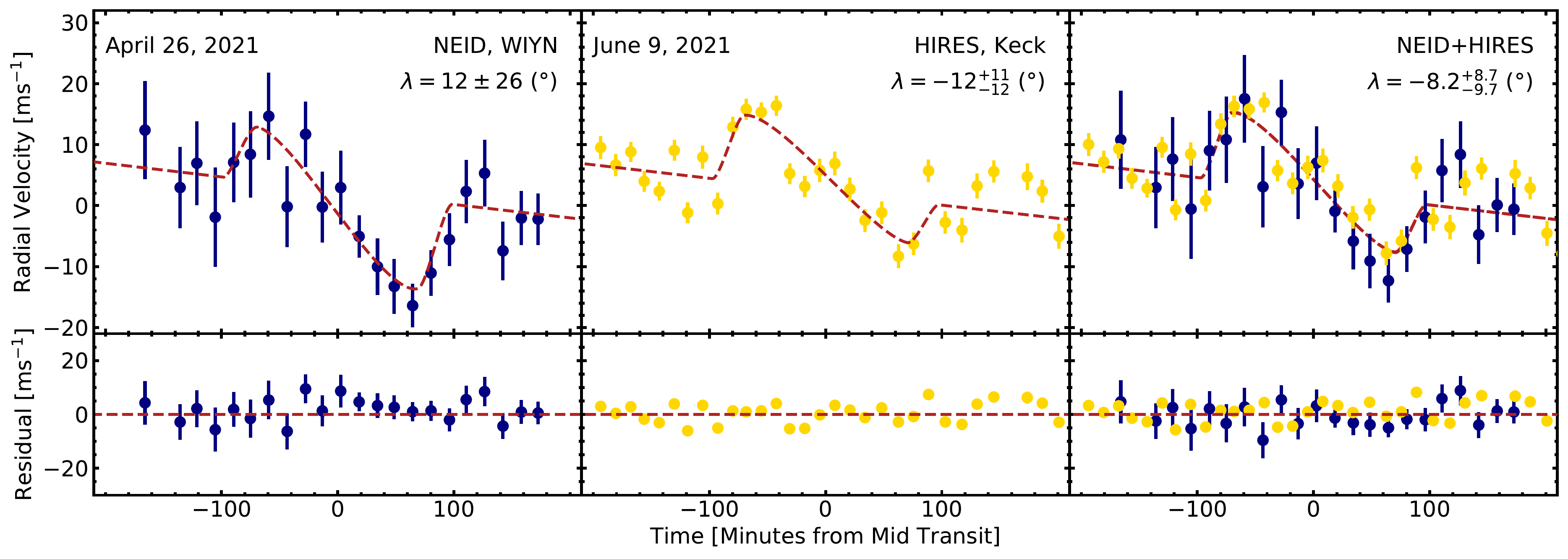}
    \caption{Spectroscopic radial velocities of WASP-148 measured with NEID (left), HIRES (center) and both together (right), as a function of orbital phase (minutes from mid-transit) along with the best-fitting R-M models shown as red dashed lines. We find a low obliquity for WASP-148b in each of the three fits. \texttt{Allesfitter} simultaneously fits the error scaling and each dataset to weight each dataset based on its likelihood. The errors of the plotted RVs are unscaled and are drawn directly from our RV pipeline. (The data used to create this figure are \href{https://github.com/wangxianyu7/Data_and_code}{available}.)}
    \label{fig:RM_fits}
\end{figure*}

\subsection{Rossiter-Mclaughlin Effect Measurements}
\subsubsection{Doppler Velocimetry with WIYN/NEID}

We observed WASP-148 with the high-resolution (R$\sim$110,000) WIYN/NEID spectrograph \citep{schwab2016design} on April 26, 2021 and obtained 24 RV measurements with fixed, 900s exposure times from UT 5:07-11:40. The night began with some high cirrus that began to clear about 30\,minutes into our observations. Seeing ranged from $0.9\arcsec-1.3\arcsec$, with airmass spanning $z=1.0-2.0$. The typical signal-to-noise ratio (S/N) is 16 pixel$^{-1}$ for the WIYN/NEID spectra at 5500 \AA.

Because WASP-148 is relatively faint compared to the NEID Fabry-P\'erot etalon calibrator, we did not take simultaneous calibrations. Instead, we obtained a set of two etalon frames both immediately prior to and following our R-M observations, in addition to NEID's standard morning and afternoon wavelength calibration sequences which include ThAr lamp, etalon, and Laser Frequency Comb (LFC) data. We also interrupted our R-M observations from UT6:03-6:16 (during the pre-transit baseline) to obtain a set of intermediate-calibration frames, including an etalon frame and three LFC frames. 

The NEID data were reduced using the NEID Data Reduction Pipeline\footnote{\leftline{More information can be found here:} https://neid.ipac.caltech.edu/docs/NEID-DRP/}, and the Level-2 1D extracted spectra were retrieved from the NExScI NEID Archive\footnote{https://neid.ipac.caltech.edu/}. We applied a modified version of the SERVAL (SpEctrum Radial Velocity anaLyzer) code \citep{zechmeister2018} to extract precise RVs using the template-matching method. This version of the code, described in \citep{stefansson2022}, is built on both the original SERVAL code and the customized version developed for the Habitable-zone Planet Finder spectrograph \citep{stefansson2020}. The code produces similar RVs to the official NEID pipeline, but with slightly smaller uncertainties (median RV errors are 5.6m/s and 5.4m/s, respectively) as we incorporate the full wavelength range of NEID. The R-M measurement of WASP-148b from NEID is shown in the middle panel of Figure~\ref{fig:RM_fits}.

For the RV extraction, we used 85 echelle orders spanning wavelengths from 398 to 895nm, masking telluric lines following \citep{stefansson2022}. Although NEID is sensitive down to 380nm, the bluer orders had low S/N and their inclusion did not improve the RV uncertainties. We used all available observations to create a master RV template. Since the moon was bright and high in the sky, we experimented with using the sky fiber to subtract the background sky from the science fiber. Doing so did not significantly change the RVs. As we obtained a slightly higher RV precision without performing the sky subtraction, we elected to extract the RVs from non-sky-subtracted spectra.

\subsubsection{Doppler Velocimetry with Keck/HIRES}

We also obtained 38 RV measurements of WASP-148 with continuous Keck/HIRES observations from UT 6:14-14:38 on June 9 2021 (dark night). Seeing ranged from $0.9\arcsec-1.1\arcsec$ during the pre-transit baseline and in-transit observations, rising slightly to $1.1\arcsec-1.4\arcsec$ during the post-transit baseline. 

\begin{figure*}
   \includegraphics[width = 1.0\textwidth]{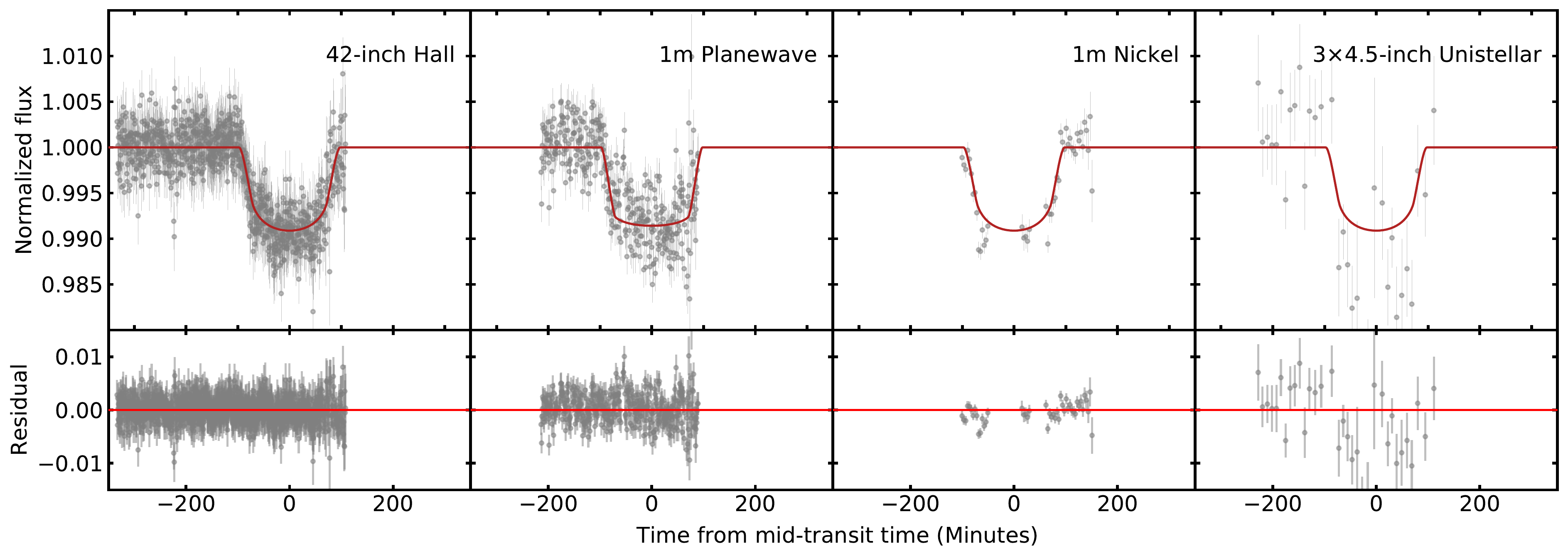}
    \caption{Photometric transit observations of WASP-148, obtained while simultaneously measuring the R-M effect using Keck/HIRES. In the upper panels, the model is plotted in red, while data is shown as gray dots. The residuals are provided in the lower panels. We find no significant spot-crossing signal, which would be indicative of high surface activity, during the transit. (The data used to create this figure are \href{https://github.com/wangxianyu7/Data_and_code}{available}.)}
    \label{fig:Transit0608}
\end{figure*}

All Keck/HIRES observations were obtained with the C2 decker ($14\arcsec\times0.861\arcsec, R=60,000$). Observations were imprinted with molecular iodine features using an iodine absorption cell to enable precise Doppler measurements \citep{Butler1996}. The median exposure time was $780\,$s, with $\sim$40k exposure-meter counts per spectrum during the first half of observations, falling to $\sim$34k exposure-meter counts per spectrum during the second half due to light cirrus. The typical S/N is $76\,$ pixel$^{-1}$ for the Keck/HIRES spectra at 5300 \AA.

During Jun. 11, we also obtained a 45-minute iodine-free HIRES template spectrum of WASP-148 using the B3 decker ($14.0\arcsec\times0.574\arcsec, R=72,000$), with $1.0\arcsec$ seeing and airmass $z=1.1$. We used this observation to calibrate our RVs with the California Planet Search pipeline  \citep{howard2010california} and to extract precise stellar parameters (see Appendix \ref{section:stellar_parameters}). At 5300 \AA, our reduced template had a S/N of 167 pixel$^{-1}$ (178k exposure meter counts). The R-M observation of WASP148b measured with HIRES is shown in the left panel of Figure~\ref{fig:RM_fits}.

\subsection{Simultaneous Photometric Observations}

The transit mid-time of WASP-148b varies by up to $\sim20\,$min due to dynamical interactions with the companion planet WASP-148c. As a result, the exact transit mid-time of WASP-148b, which helps to better constrain the R-M model, cannot be predicted from past observations as accurately as in most systems.

To directly measure the transit mid-time, we took photometric observations of WASP-148 using the 42-inch Hall telescope and 1-m Planewave telescope at Lowell Observatory; the 1-m Nickel telescope at Lick Observatory; and an array of seven 4.5-inch Unistellar eVscopes located in California and North Carolina while simultaneously observing the R-M effect with Keck/HIRES. The obtained light curves are provided in Figure \ref{fig:Transit0608}. Observing and data reduction details are provided in Appendix \ref{photometry_appendix}.

In addition to the simultaneous photometric observations, our analysis also incorporates two photometric transits obtained with the 1.5-m Ritchey-Chr\'{e}tien Telescope \citep{Maciejewski2020}, as well as \textit{TESS} light curves derived from the MIT QuickLook Pipeline \citep{Huang2020QLP}. Seven transits of WASP-148b were observed in \textit{TESS} Sectors 24-26 from April 16 to July 04, with $30\,$-minute exposures. To remove potential trends, we employed the Savitzky-Golay filter over a 12-hour window to the light curves.

\section{Stellar Obliquity from Global Analysis}

We determined the sky-projected spin-orbit angle ($\lambda$) for WASP-148b using the \texttt{Allesfitter} code (\citealt{allesfitter-code}). \texttt{Allesfitter} simultaneously fits multi-band transits, RV data, and R-M measurements by applying the affine-Invariant Markov Chain Monte Carlo (MCMC) algorithm.

We simultaneously modeled 11 photometric transits (seven from \textit{TESS}, two from the 1.5-m Ritchey-Chr\'{e}tien Telescope \citep{Maciejewski2020}, and two newly collected light curves from the 42-inch Hall Telescope and the 1-m Planewave Telescope), the in-transit NEID and HIRES RVs (which exhibit the R-M effect), and the out-of-transit RVs available from the system's discovery paper \citep{hebrard2020discovery}, taken with the SOPHIE spectrographs.

The model parameters include the orbital period ($P$), transit mid-time at a reference epoch ($T_{0}$), mid-time for each transit event ($T_{n}$), cosine of the orbital inclination ($\cos{i}$), planet-to-star radius ratio ($R_{P}/R_{\star}$), sum of radii divided by the orbital semi-major axis ($(R_{\star}+R_{P})/a$), RV semi-amplitude ($K$), parameterized eccentricity and argument of periastron ($\sqrt{e}\,\cos{\,\omega}$, $\sqrt{e}\,\sin{\,\omega}$), transformed quadratic limb-darkening coefficients ($q_{1}$ and $q_{2}$)\footnote{The relationship between transformed and physical quadratic limb-darkening coefficients is given as $u_{1}=2 q_{2}  \sqrt{q_{1}}$ and $u_{2}=\sqrt{q_{1}}(1-2 q_{2})$ as described in \cite{Max2021}.} , jitter term (ln$\sigma_{\rm jitter}$), sky-projected spin-orbit angle ($\lambda$), and sky-projected stellar rotational velocity ($v\sin i_{\star}$). All of the listed parameters were fitted for WASP-148b, while only $P$, $K$, $\sqrt{e}\,\cos{\,\omega}$, and $\sqrt{e}\,\sin{\,\omega}$ were fitted for WASP-148c, since it does not transit.

Uniform priors were adopted for all fitted parameters. Initial guesses for them were set to the values reported in \citet{hebrard2020discovery}. Each Rossiter-McLaughlin fit includes an additive offset that jointly accounts for instrumental systematics and stellar variability, determined through a second-order hybrid polynomial model added on to the R-M fit \citep{Max2021}. Table \ref{table:results} summarizes the model parameters and priors.

\begin{deluxetable*}{lccccc}
\tablecaption{System Parameters, Priors, and Results for WASP-148  \label{table:results}}
\tabletypesize{\scriptsize}
\tablehead{
  \colhead{ } & \colhead{HIRES Spectrum}  & \colhead{MIST+SED} &\colhead{ }&\colhead{ }      \\
 \colhead{} & \colhead{The Cannon}  & \colhead{EXOFASTv2}  & \colhead{ }&\colhead{ }
}
\tablewidth{300pt}
\startdata
\multicolumn{5}{l}{Stellar Parameters:}\\
~~~~$M_*$ (\msun) & - & $0.970^{+0.046}_{-0.057}$ &&& \\
~~~~$R_*$ (\rsun) & - & $0.905^{+0.013}_{-0.014}$ &&&  \\
~~~~$\log{g}$ (cgs) & $4.48\pm0.10$ & $4.511^{+0.023}_{-0.028}$ &&&  \\
~~~~$[\rm{Fe/H}]$ (dex) & $0.25\pm0.04$ & $0.35^{+0.14}_{-0.19}$ && & \\
~~~~$T_{\rm eff}$ (K) & $5478\pm59$ & $5437\pm21$ &&&  \\
~~~~$v\sin i_{\star}$ (km/s) & $2.44\pm1.07$ &- &&&  \\
\hline
\hline
  &Priors for global fit&Global fit 1: NEID&Global fit 2: HIRES &Global fit 3: NEID+HIRES  \\
  & & & &(Preferred Solution)  \\
\hline
\multicolumn{5}{l}{Stellar Parameters:}\\
~~~~$v\sin i_{\star}$ ($\rm km \ s^{-1}$)          &  $\mathcal U(2.44; 0.0; 10)$                & $2.95_{-0.67}^{+0.74}$            &  $2.11_{-0.37}^{+0.39}$        &$2.30_{-0.34}^{+0.38}$             \\  
\multicolumn{5}{l}{Planetary Parameters:}\\
\multicolumn{5}{l}{~~~~\textbf{WASP-148b:}}\\
~~~~$\lambda_{b}$ (deg)                            &  $\mathcal U(0; -180; +180)$                 & $12\pm26$                        &  $-12_{-12}^{+11}$             &$-8.2_{-9.7}^{+8.7}$         \\                                    
~~~~$P_{\rm b}$ (days)                             &  $\mathcal U(8.8035; 8.0; 10.0)$             & $8.803569\pm0.000022$       &  $8.803554\pm0.000021$       &\textbf{$8.803544\pm0.000021$}  \\                           
~~~~$R_{P;b} \ (\rj)$                              & -                              					    & $0.807_{-0.020}^{+0.018}$        &  $0.799_{-0.022}^{+0.020}$     &$0.800_{-0.023}^{+0.020}$     \\                                                                         
~~~~$M_{P;b} \ (\mj)$                              & -                              					    & $0.368_{-0.049}^{+0.053}$        &  $0.352_{-0.050}^{+0.056}$      &$0.354_{-0.050}^{+0.055}$         \\ 
~~~~$T_{0;b} \ (\rm BJD_{\rm TDB} - 2450000) $     &    $\mathcal U(9163.62010;9163.0; 9164.0)$ 	& $9163.62079\pm0.00037$         &  $9163.62042\pm0.00036$          &$9163.62033\pm0.00036$               \\  
~~~~$i_{b}$ (deg)                                  & -                              					    & $86.88_{-0.78}^{+0.68}$          &  $87.17_{-0.81}^{+0.73}$         &$87.14_{-0.82}^{+0.75}$                    \\                                                                        
~~~~$e_{b}$                                        & -                              					    & $0.190\pm0.075$                   &  $0.187\pm0.077$               &$0.190_{-0.072}^{+0.078}$                   \\                                                                        
~~~~$\omega_{b}$ (deg)                             & -                              					    & $66_{-18}^{+14}$                  &  $63_{-31}^{+16}$              &$65_{-21}^{+13}$                  \\       
~~~~$\cos{i_b}$                                    &  $\mathcal U(0.0350; 0.0; 1.0)$              & $0.054_{-0.012}^{+0.013}$         &  $0.049_{-0.013}^{+0.014}$               &$0.050_{-0.013}^{+0.014}$               \\                              
~~~~$K_{\rm b}$ ($\rm m \ s^{-1}$)                 &  $\mathcal U(29.5; 25.0; 35.0)$              & $28.9\pm2.0$                      &  $28.8\pm2.0$                   &$ 28.8\pm2.0$             \\                                
~~~~$R_P / R_\star$                                &  $\mathcal U(0.0950; 0.0; 1.0)$              & $0.0918_{-0.0018}^{+0.0015}$      &  $0.0909_{-0.0021}^{+0.0017}$    &$0.0909_{-0.0022}^{+0.0017}$  \\                                          
~~~~$(R_\star + R_b) / a_b$                        &  $\mathcal U(0.0638; 0.0; 1.0)$              & $0.0732\pm0.0079$                 &  $0.0699\pm0.0084$                &$0.0703\pm0.0084$             \\  
~~~~$\sqrt{e_b} \cos{\omega_b}$                    &  $\mathcal U(0.2547; -1.0; 1.0)$             & $0.178_{-0.096}^{+0.090}$         &  $0.17_{-0.12}^{+0.11}$            &$0.171_{-0.093}^{+0.086}$               \\                              
~~~~$\sqrt{e_b} \sin{\omega_b}$                    &  $\mathcal U(0.3296; -1.0; 1.0)$             & $0.388_{-0.13}^{+0.097}$          &  $0.38_{-0.14}^{+0.10}$         &$0.394_{-0.13}^{+0.097}$     \\            
\multicolumn{5}{l}{~~~~\textbf{WASP-148c:}}\\
~~~~$P_{\rm c}$ (days)                            &  $\mathcal U(34.5160; 30.0; 40.0)$      	  & $34.527\pm0.024$                  &  $34.526_{-0.025}^{+0.023}$           &$34.527\pm0.024$                 \\  
~~~~${\emph{M}_{P;c} \mathrm{sin} i \ (\mj)}$                           & -                              					    & ${0.407_{-0.088}^{+0.127}}$        &  ${0.404_{-0.090}^{+0.131}}$      &${0.408_{-0.087}^{+0.127}}$         \\ 
~~~~$e_{c}$                                       & -                                         	& $0.350\pm0.064$                   &$0.349_{-0.066}^{+0.062}$              &$0.351_{-0.064}^{+0.060}$           \\                                                             
~~~~$\omega_{c}$ (deg)                            & -                                         	& $12_{-13}^{+14}$                  &$13_{-14}^{+15}$             &$12_{-13}^{+14}$                  \\                          
~~~~$T_{0;c} \ (\rm BJD_{\rm TDB} - 2450000) $    &  $\mathcal U(8073.264; 8073.0; 8074.0)$     & $8073.34_{-0.68}^{+0.61}$         &  $8073.34_{-0.69}^{+0.62}$  &$ 8073.31_{-0.67}^{+0.63}$        \\ 
~~~~$K_{\rm c}$ ($\rm m \ s^{-1}$)              &  $\mathcal U(25.9; 20.0; 30.0)$       	      & $27.1_{-2.6}^{+2.8}$              &  $26.9\pm2.7$               &$ 27.2\pm2.7$                \\                     
~~~~$\sqrt{e_c} \cos{\omega_c}$                 &  $\mathcal U(0.58; -1; 1.0)$              	  & $0.568_{-0.076}^{+0.058}$         &  $0.564_{-0.079}^{+0.060}$  &$0.568_{-0.075}^{+0.058}$        \\ 
~~~~$\sqrt{e_c} \sin{\omega_c}$                 &  $\mathcal U(0.14; -1; 1.0)$              	  & $0.12\pm0.13$                     &  $0.13\pm0.14$              &$0.12\pm0.13$         \\
~~~~  $ \ln{\sigma_\mathrm{jitter; NEID}}$ ($\ln{ \mathrm{km/s} }$)                 &  ${ \mathcal U(-3; -15; 0)}$              	  & ${-10.6\pm3.0}$                     & -              &${-10.7\pm3.0}$        \\
~~~~$\ln{\sigma_\mathrm{jitter; HIRES}}$ ($\ln{ \mathrm{km/s} }$)                &   ${\mathcal U(-3; -15; 0)}$              	  & -                     &  ${-5.70\pm0.16}$              &${-5.70\pm0.16}$          \\
\multicolumn{5}{l}{Transit Mid-times for WASP-148b:}\\
~~~~$T_{1}\ (\rm BJD_{\rm TDB} - 2450000) $      &  $\mathcal N(8961.12366; 0.1)$               & $8961.1237_{-0.0027}^{+0.0025}$                &  $8961.1238_{-0.0026}^{+0.0024}$             &$ 8961.1238_{-0.0026}^{+0.0024}$          \\
~~~~$T_{2}\ (\rm BJD_{\rm TDB} - 2450000) $      &  $\mathcal N(8969.92904; 0.1)$               & $8969.9291\pm0.0029$                           &  $8969.9291_{-0.0030}^{+0.0028}$             &$ 8969.9291\pm0.0030$          \\
~~~~$T_{3}\ (\rm BJD_{\rm TDB} - 2450000) $      &  $\mathcal N(8978.73498; 0.1)$               & $8978.7347_{-0.0020}^{+0.0022}$                &  $8978.7347_{-0.0020}^{+0.0021}$             &$ 8978.7347_{-0.0020}^{+0.0021}$          \\
~~~~$T_{4}\ (\rm BJD_{\rm TDB} - 2450000) $      &  $\mathcal N(8987.54213; 0.1)$               & $8987.54200_{-0.00071}^{+0.00066}$             &  $8987.54199_{-0.00073}^{+0.00065}$          &$ 8987.54197_{-0.00073}^{+0.00066}$       \\
~~~~$T_{5}\ (\rm BJD_{\rm TDB} - 2450000) $      &  $\mathcal N(9005.15336; 0.1)$               & $9005.15315_{-0.00100}^{+0.0011}$              &  $9005.15304_{-0.00096}^{+0.0010}$           &$ 9005.15303_{-0.00094}^{+0.0010}$        \\
~~~~$T_{6}\ (\rm BJD_{\rm TDB} - 2450000) $      &  $\mathcal N(9013.95526; 0.1)$               & $9013.9557_{-0.0054}^{+0.0045}$                &  $9013.9557_{-0.0055}^{+0.0044}$             &$ 9013.9556_{-0.0055}^{+0.0044}$          \\
~~~~$T_{7}\ (\rm BJD_{\rm TDB} - 2450000) $      &  $\mathcal N(9031.57093; 0.1)$               & $9031.5707_{-0.0026}^{+0.0028}$                &  $9031.5705_{-0.0025}^{+0.0028}$             &$ 9031.5704_{-0.0025}^{+0.0028}$          \\
~~~~$T_{8}\ (\rm BJD_{\rm TDB} - 2450000) $      &  $\mathcal N(9040.37324; 0.1)$               & $9040.3731_{-0.0014}^{+0.0013}$                &  $9040.3736_{-0.0013}^{+0.0011}$             &$ 9040.3736_{-0.0013}^{+0.0012}$          \\
~~~~$T_{9}\ (\rm BJD_{\rm TDB} - 2450000) $      &  $\mathcal N(9084.39824; 0.1)$               & $9084.39808\pm0.00056$                         &  $9084.39800\pm0.00057$                      &$ 9084.39800\pm0.00056$                   \\
~~~~$T_{10}\ (\rm BJD_{\rm TDB} - 2450000) $     &  $\mathcal N(9330.87464; 0.1)$               & $9330.8684_{-0.0061}^{+0.0065}$                &  -                                           &$9330.8651_{-0.0045}^{+0.0048}$           \\
~~~~$T_{11}\ (\rm BJD_{\rm TDB} - 2450000) $     &  $\mathcal N(9374.90075; 0.1)$               & $9374.90171\pm0.00081$                         &  $9374.90148_{-0.00074}^{+0.00079}$          &$9374.90149\pm0.00077$                   \\
\multicolumn{5}{l}{Transformed limb darkening coefficients:}\\
~~~~$q_{\rm 1:NEID}$     &  $\mathcal U(0;0.5;1)$               & $0.47_{-0.30}^{+0.25}$                         &  $-$          & $0.39_{-0.26}^{+0.30}$                  \\
~~~~$q_{\rm 2:NEID}$     &  $\mathcal U(0;0.5;1)$               & $0.36\pm0.24$                         &  $-$          &$0.33_{-0.23}^{+0.25}$                   \\
~~~~$q_{\rm 1:HIRES}$     &  $\mathcal U(0;0.5;1)$               & $-$                         &  $0.37_{-0.25}^{+0.30}$          &$0.42\pm0.27$                   \\
~~~~$q_{\rm 2:HIRES}$     &  $\mathcal U(0;0.5;1)$               & $-$                         &   ${0.33_{-0.23}^{+0.25}}$          &$0.34\pm0.24$                   \\
\multicolumn{5}{l}{Physical limb darkening coefficients:}\\
~~~~$u_{\rm 1:NEID}$     &  -               & $0.42_{-0.29}^{+0.40}$                         &  $-$          &  $0.34_{-0.24}^{+0.37}$                 \\
~~~~$u_{\rm 2:NEID}$     &  -               & $0.15_{-0.26}^{+0.35}$                         &  $-$          &$0.17_{-0.24}^{+0.33}$                  \\
~~~~$u_{\rm 1:HIRES}$     &  -               & $-$                         &  $0.33_{-0.24}^{+0.37}$          &$0.37_{-0.26}^{+0.38}$                   \\
~~~~$u_{\rm 2:HIRES}$     &  -               & $-$                         &   $0.16_{-0.24}^{+0.33}$          &$0.17_{-0.25}^{+0.33}$                  \\
\enddata 
\tablenotetext{}{\, \, Parameter definitions are provided in Section 3. $i$, $e$, and $\omega$ are derived parameters with no priors; all other parameters with priors are directly fitted. $T_{1}-T_{11}$ are the mid-times for each transit event. There is no $T_{10}$ from the HIRES global fit because this measurement is derived from the NEID R-M observation, with no joint photometry to constrain the transit mid-time on that night.}
\end{deluxetable*}

We conducted three joint fits. First, we fit two separate models to independently obtain $\lambda$ from each of the two in-transit R-M measurements collected by WIYN/NEID and Keck/HIRES, respectively. Then, we modeled the combined dataset. Together with the in-transit RVs, each fit incorporated the \textit{TESS} and ground-based telescope photometry, as well as the out-of-transit RV data. 

For each fit, we sampled the posterior distributions of the model parameters using the Affine-Invariant MCMC algorithm with 100 independent walkers each with 200{,}000 total accepted steps. All Markov chains were run to over 30$\times$ their autocorrelation lengths, such that they were fully converged. The results, listed in Table~\ref{table:results}, include solutions for transit and RV parameters, as well as $\lambda_{\rm b}$, $v\sin i_{\star}$, and the associated $1\sigma$ uncertainties. The best-fitting model for each R-M observation is shown in Figure \ref{fig:RM_fits}. 

The RV and transit parameters obtained from our analysis are in good agreement with the values from \citet{hebrard2020discovery}.
We find the best-fit $\lambda$ and $v\sin i_{\star}$ from the phased transit to be $\lambda = -8\degree.2^{+8\fdeg.7}_{-9\fdeg.7}\,$ and $v\sin i_{\star} = 2.30^{+0.38}_{-0.34}$\,\kms, in perfect agreement with the corresponding values from independent fits to the two transit events. Our results suggest that the orbit of WASP-148b is aligned with the spin axis of its host star.

\section{Discussion}

\begin{figure}
   \includegraphics[width = 1.0\columnwidth]{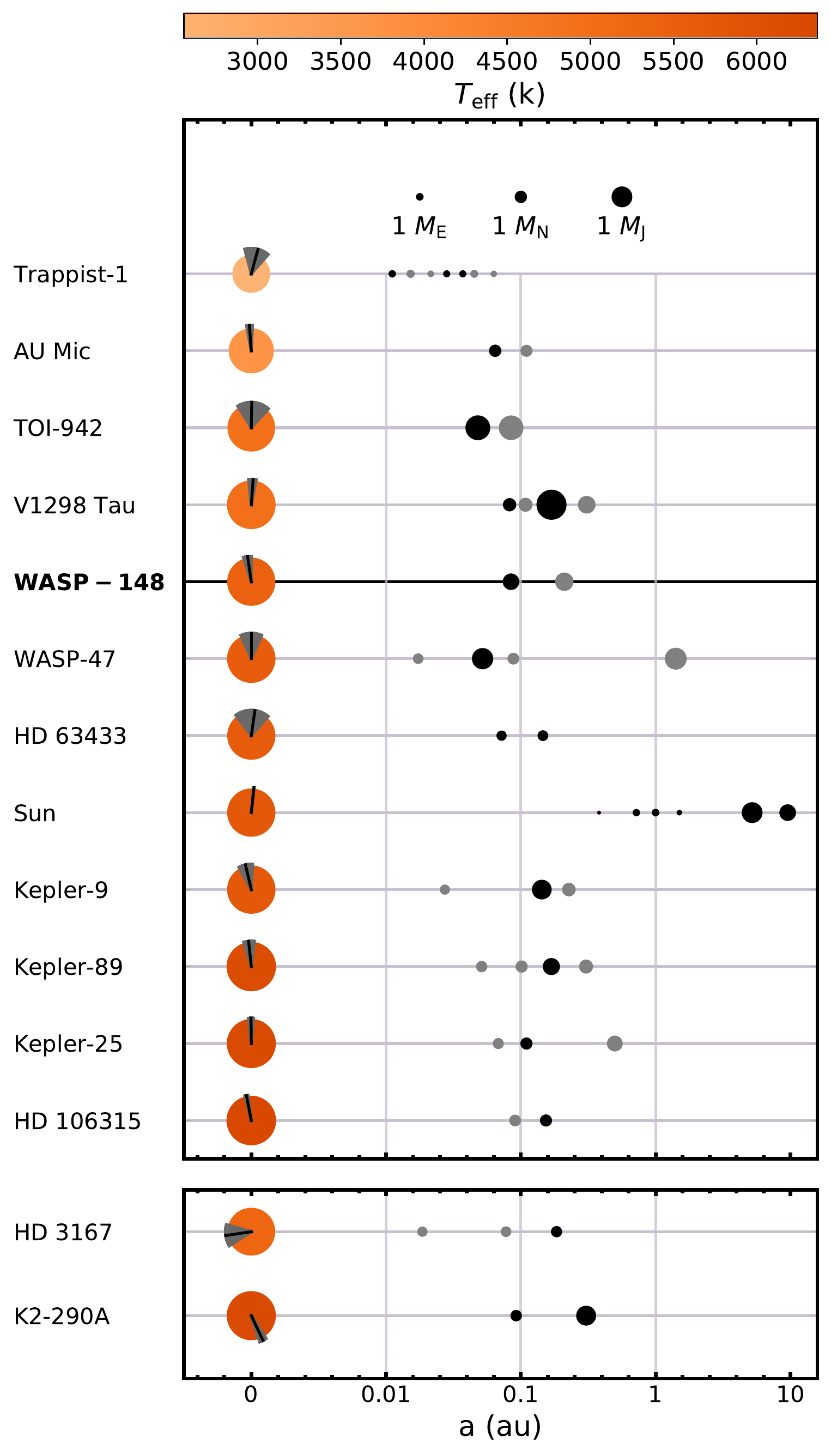}
    \caption{Compact multi-planet systems with observed $\lambda$ constraints, ordered by stellar temperature. For each system, $\lambda$ is shown, together with uncertainties, relative to the net angular momentum plane of the orbits. This plane is determined from those planets with measured spin-orbit angles in each system. A vertical upwards spin axis indicates $\lambda=0\degree$. The planets with spin-orbit angle constraints are shown in black, while all other planets are shown in gray. The planetary masses in the TOI-942 and V1298 Tau systems were estimated based on their photometric radii using the \citet{Chen2017} mass-radius relation.}
    \label{fig:multiplanet_systems}
\end{figure}

WASP-148 is the fourteenth compact multi-planet system with a measured $\lambda$. We define ``compact'' systems as those with small period ratios (within our sample, $P_2/P_1 < 6$) in which neighboring planets actively dynamically interact such that the systems behave as a whole. Figure \ref{fig:multiplanet_systems} provides an overview of these systems, along with the solar system for reference. We exclude the 55\,Cancri system due to its contested obliquity measurement \citep{lopez2014rossiter}.

All extrasolar systems in Figure \ref{fig:multiplanet_systems} other than WASP-148 are multi-transiting systems. Because the most updated available information suggests a low mutual inclination between the WASP-148 planets \citep{Maciejewski2020}, we treat WASP-148 as an analog to the multi-transiting systems. We note that multi-transiting systems are also not guaranteed to be coplanar, since mutually inclined planets can transit the host star with different trajectories.

Most compact multi-planet systems are consistent with alignment, as shown by the top panel of Figure \ref{fig:multiplanet_systems}.
This finding may point to an early history in which migration and accretion occurred in isolation, with relatively little disturbance. The absence of dynamically violent interactions is consistent with the presence of multiple planets with small observed mutual inclinations in these systems. 

We define a ``misaligned'' system as a system with a $\lambda$ that exceeds 10\degree\, and differs from 0\degree\, at the 3$\sigma$ level. The two misaligned compact multi-planet systems -- HD\,3167 and K2-290A -- are separated out from the aligned systems in Figure \ref{fig:multiplanet_systems}. The residuals of the R-M fit to the HD 3167 system \citep{Dalal2019}, however, have an amplitude comparable to the R-M signal itself ($\sim$ 1m s$^{-1}$ ). As a result, we include only K2-290A \citep{Hjorth2021}, hosting two transiting planets, in our population analysis. We note that the misalignment of K2-290A could be produced by the companion stars in the system, rather than due to dynamically violent interactions within the observed planetary system.

In our population analysis, we included only systems with $\lambda$ determined through the R-M effect or Doppler Tomography. This choice was made for two reasons: (1) to provide a clean sample with a single set of detection biases, and (2) to constrain our sample to only systems around main-sequence stars, since it is not well understood how stellar evolution should alter a system's spin-orbit angle. We note that the inclusion of misaligned systems observed with separate methods around post-main-sequence stars, such as Kepler-56 \citep{Huber2013} and Kepler-129 \citep{zhang2021long}, would further weaken the modest statistical significance, characterized in Section \ref{subsection:statistical_significance}, of the trend towards alignment for compact multi-planet systems.

\subsection{Statistical Significance of the Aligned Compact Multi-planet System Trend}
\label{subsection:statistical_significance}
Only one out of twelve compact multi-planet systems with $\lambda$ measurements is unambiguously misaligned. In the following analyses, we examine whether this sample is large enough to determine whether the obliquity distribution between isolated-hot-Jupiter and compact multi-planet systems is significantly different. We find that the difference between these two populations is only statistically significant ($3.7\sigma$) after systems that may have been tidally realigned are removed. In contrast, a direct or temperature-controlled comparison between isolated-hot-Jupiter and compact multi-planet systems does not produce a significant difference.

\begin{figure*}
   \includegraphics[width = 1.0\textwidth]{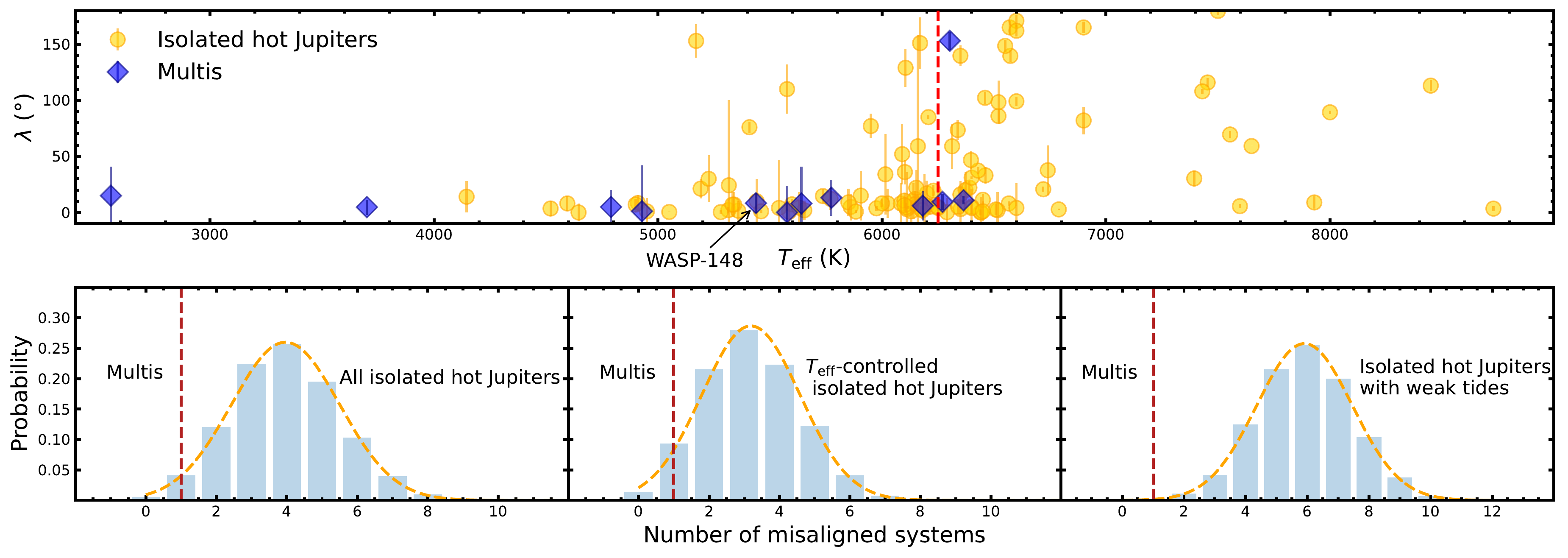}
    \caption{$Upper \ panel$: Stellar sky-projected obliquities ($\lambda$) derived from R-M measurements or Doppler Tomography for 112 exoplanet systems, shown as a function of the host star's effective temperature ($T_{\rm eff}$). The isolated-hot-Jupiter systems are shown as yellow circles, while compact multi-planet systems are represented by blue diamonds. The vertical red dashed line indicates $\teff= 6,250\,$K.\\
    $Lower \ panel$: The probability distribution for the number of the misaligned systems among 12 randomly selected isolated-hot-Jupiter systems, shown for each of our three test cases. In each panel, the dashed red line represents only one misaligned system, K2-290A, among 12 compact multi-planet systems. Singles with weak tides include systems with low-mass planets, wide-separation systems, or systems around hot stars.}
    \label{fig:distribution_probability}
\end{figure*}

\textbf{\textit{Isolated hot Jupiters vs. Compact Multis.}} 
Randomly drawing 12 systems from the $112$ isolated hot-Jupiter systems with $\lambda$ measurements, there is a $4.1\%$ chance that one of them is misaligned (lower left panel of Figure~\ref{fig:distribution_probability}). Compact multi-planet systems are preferentially more aligned than isolated-hot-Jupiter systems at the $2.5\,\sigma$ level.

\textbf{\textit{Correlation with Stellar Temperature.}}
Previous population studies have provided evidence that hot ($\teff > 6,250\,$K) and high-mass ($M_{*} > 1.1 \ \msun$) stars tend to have high obliquities \citep{Schlaufman2010, Winn2010, Albrecht2012}. We therefore further test the difference in obliquity distributions between isolated-hot-Jupiter and compact multi-planet systems with a \teff-controlled sample. We randomly draw 12 systems from $83$ isolated-hot-Jupiter systems that have a range of host star temperatures between 2,557\,K and 6,364\,K, which is the same as that of the compact multi-planet system sample. We find a $9.5\%$ probability that one out of 12 randomly selected isolated-hot-Jupiter systems from the \teff-controlled sample is misaligned. After controlling for stellar \teff, the significance of the difference between isolated-hot-Jupiter and compact multi-planet system obliquities drops to $2.2\sigma$, as shown in the lower central panel of Figure \ref{fig:distribution_probability}.

All ten compact multi-planet systems with host star temperature $\teff< 6,250$\,K and $\lambda$ measurements are aligned. It is unclear whether this trend arises because compact multi-planet systems are generally spin-orbit aligned or because cool stars with $\teff < 6,250$\,K generally have low obliquities. So far, there are only three $\lambda$ measurements for compact multi-planet systems orbiting host stars with $\teff > 6,250$\,K. Interestingly, one of them, K2-290A with $M_*=1.2{\rm M_\sun}$ and $T_{\rm eff}=6,301\,$K, is retrograde (as shown in the upper panel of Figure \ref{fig:distribution_probability}). 

Based on the small existing sample of compact multi-planet systems, the ${\teff-\lambda}$ relation observed for isolated-hot-Jupiter systems appears to also hold for compact multi-planet systems. R-M measurements for compact multi-planet systems, especially those around hot stars, are urgently needed to confirm this trend with a larger sample.

\textbf{\textit{The Influence of Tides.}}
The ${\teff-\lambda}$ relation is generally viewed as evidence for tidal dissipation, which depends sensitively on the internal structure of the host star and is a strong function of orbital separation and planet mass. This idea is supported by the previously observed trend that the high-mass ($M_{\rm}>0.3M_{\rm J}$), closest-orbiting planets ($a/R_*<12$) around cool stars ($\teff<6,250$K) tend to have well-aligned orbits, suggesting that these systems were realigned through tidal dissipation \citep{Wang2021}. We therefore exclude isolated-hot-Jupiter systems with $a/R_*<12$ and $\teff<6,250$K, which may have had their misalignments erased by tides, before comparing them to compact multi-planet systems which were unlikely to be influenced by tides because of their relatively long orbital periods and low planet masses.

We randomly draw 12 systems from the sample of 55 isolated-hot-Jupiter systems with weak tides. We found that there is only a $0.17\%$ probability that one out of 12 randomly selected systems is misaligned. Based on this test, the compact multi-planet systems are preferentially more aligned than the isolated-hot-Jupiter systems at 3.7$\sigma$ significance.

Under the tidal realignment assumption, the difference in obliquity distribution between isolated-hot-Jupiter and compact multi-planet systems is significant. The tidal realignment hypothesis, however, has an outstanding theoretical issue: it is unclear how a planet can re-align the host star's spin axis without sacrificing all of its angular momentum and inspiraling into the host star (see \citealt{Anderson2021} for the recent progress on this problem).

Most of the compact multi-planet systems in our sample orbit cool stars, while the isolated-hot-Jupiter systems retained in this analysis orbit mostly hot stars. Therefore, this result may be indicative of the difference in obliquities across stellar temperatures \citep[e.g.][]{Winn2010}, rather than a systematic difference between isolated-hot-Jupiter and multi-planet systems.

\subsection{Implications for the Origins of Spin-Orbit Misalignments}

In addition to individual R-M measurements, statistical arguments have been used to investigate the typical obliquity of planetary hosts based on the $v\sin i_{\star}$ method. From these studies, it is clear that hot and massive stars harboring hot Jupiters are more commonly misaligned \citep{Schlaufman2010}, in agreement with the results from R-M measurements \citep{Winn2010}. It is not yet clear, however, whether this trend arises because all hot stars generally tend to have high obliquities \citep{Louden2021} or instead because hot Jupiter host stars are more likely to be misaligned \citep{Winn2017}. Some candidate misaligned systems with small planets have been identified \citep{Walkowicz2013, Hirano2014, Morton2014}, and one, K2-290 A, has been confirmed \citep{Hjorth2021}, implying that the former may be true.

Distant giant planets are more common around high-mass and hot stars, which are also more likely to host a wide range of misaligned planets: not only single hot Jupiters, but also compact systems with multiple super-Earths \citep{Louden2021, Hjorth2021}. As an alternative to the theory of tidal dissipation, the excess of misaligned systems around hot stars suggests that spin-orbit misalignments may be caused by distant giant perturbers, which have been found to be common in systems that already host close-in planets \citep{Ngo2015, Bryan2019, Zhu2018SE_CJ}.

The distant giant perturbers may sometimes tilt coplanar planet systems as a whole, such as in the Kepler-56 system \citep[e.g.][]{Li2014,Gratia2017}. As the spin-orbit coupling timescale between the stellar $J_2$ and the inner planets is generally longer than the secular timescale of an inner planet system with distant giant perturbers, this results in the generation of mild spin-orbit misalignment angles comparable to the misalignment angles between the distant perturbers and inner compact multi-planet systems (of order $\sim 12^{\circ}$, see \citealt{Lai2018} and \citealt{masuda}).

Alternatively, the distant giant perturbers may excite not only high obliquities but also large mutual inclinations, reducing the number of simultaneously transiting planets \citep{Pu2021}. The observed hint at a correlation between stellar obliquities and planetary loneliness may, in this case, actually be a correlation between stellar obliquities and planetary mutual inclinations.

This connection between spin-orbit misalignments and distant giant perturbers will be tested by ongoing long time-baseline RV campaigns and upcoming planet search surveys–e.g., the Gaia and Nancy Grace Roman missions, which are expected to detect such exo-Jupiters in large numbers. The combination of RVs and Gaia astrometry will also provide us the 3D configuration of distant giant perturbers, which is key to understanding whether such perturbers are mutually inclined enough to tilt the orbits of inner planets. True 3D spin-orbit angle measurements for inner planet systems with known distant perturbers \citep[e.g., HAT-P-13:][]{Bakos2009, Winn2010} are also helpful to characterize the connection between distant exo-Jupiters and the \textit{true} spin-orbit misalignments of inner systems. If a robust empirical link can be established between the presence of distant giant perturbers and the misalignments of inner planet systems – as we have hypothesized and as has been observed in the HAT-P-11 \citep{Yee2018} system – then the presence of spin-orbit misalignments takes on a new meaning as an imprinted record of distant, undiscovered giant planets.

\acknowledgements

We thank the anonymous reviewer for insightful and constructive suggestions that greatly strengthened the scientific content of this paper. M.R. is supported by the National Science Foundation Graduate Research Fellowship Program under grant No. DGE-1752134. J.A.G.J acknowledges support from grant HST-GO-15955.004-A from the Space Telescope Science Institute, which is operated by the Association of Universities for Research in Astronomy, Inc., under NASA contract NAS 5-26555. P.D. is supported by a National Science Foundation (NSF) Astronomy and Astrophysics Postdoctoral Fellowship under award AST-1903811. This work was partially supported by funding from the Center for Exoplanets and Habitable Worlds. The Center for Exoplanets and Habitable Worlds is supported by the Pennsylvania State University, the Eberly College of Science, and the Pennsylvania Space Grant Consortium. This paper contains data taken with the NEID instrument, which was funded by the NASA-NSF Exoplanet Observational Research (NN-EXPLORE) partnership and built by Pennsylvania State University. NEID is installed on the WIYN telescope, which is operated by the NSF's National Optical-Infrared Astronomy Research Laboratory, and the NEID archive is operated by the NASA Exoplanet Science Institute at the California Institute of Technology. NN-EXPLORE is managed by the Jet Propulsion Laboratory, California Institute of Technology under contract with the National Aeronautics and Space Administration. This work was performed, in part, for the Jet Propulsion Laboratory, California Institute of Technology, sponsored by the United States Government under the Prime Contract 80NM0018D0004 between Caltech and NASA. The Center for Exoplanets and Habitable Worlds and the Penn State Extraterrestrial Intelligence Center are supported by the Pennsylvania State University and the Eberly College of Science. This work is supported by Astronomical Big Data Joint Research Center, co-founded by National Astronomical Observatories, Chinese Academy of Sciences and Alibaba Cloud.
\clearpage
\appendix 
\section{Photometry Observing Details and Data Reduction}
\label{photometry_appendix}

\subsection{Photometry with 42-inch Hall Telescope at Lowell Observatory}
\label{Hall42}

We obtained $15/20$s exposures continuously over $442$ min, spanning a nearly full transit of WASP-148b, on 2021 June 9 in the \textit{VR} band using the 42-inch Hall telescope at Lowell observatory. The telescope is equipped with a $ 4\mathrm{K} \times 4\mathrm{K} $ CCD camera with a pixel scale of $0.327 \arcsec \rm \, pixel^{-1}$, resulting in a field of view of $ 22.3 \arcmin \times 22.3\arcmin $. For our observations, 3$\times$3 binning mode was applied, resulting in a image-scale of 1\arcsec.05/pixel.

Standard calibrations and differential aperture photometry were implemented using AstroImageJ \citep{Collins2017}. Circular apertures with an 8-pixel ($8.4 \arcsec$) radius were applied to WASP-148 and 5 comparison stars to extract the differential light curve. We performed barycentric corrections to convert the record timestamps for each measurement to ${\rm BJD_{TDB}}$. 
 
The resulting light curve is presented in the leftmost panel of Figure~\ref{fig:Transit0608}. Since weather conditions were excellent throughout the night, with no moon or clouds in the sky, we successfully detected the transit with a photometric precision of $2\,$mmag.

\subsection{Photometry with $1$-m Planewave Telescope at Lowell Observatory}

We obtained $30$s exposures continuously over $392$ min, spanning a nearly full transit of WASP-148b, on 2021 June 9 in the Sloan \textit{r} band using the $1$-m Planewave telescope at Lowell observatory. The telescope is equipped with a $2\mathrm{K} \times 2\mathrm{K}$ CCD camera with a pixel scale of $0.621 \arcsec\, \rm{pixel}^{-1}$, resulting in a field of view of  $ 21.2\arcmin\times21.2\arcmin$. 

The images were reduced and data was extracted as described in Section~\ref{Hall42}, with 5 comparison stars and a photometric aperture of $10$ pixels ($6.21 \arcsec$). The resulting light curve is presented in the middle left panel of Figure~\ref{fig:Transit0608}. Since weather conditions were excellent throughout the night, with no moon or clouds in the sky, we successfully detected the transit with a photometric precision of $2\,$mmag .

\subsection{Photometry with $1$-m Nickel Telescope at Lick Observatory}

We observed the transit of WASP-148b on 2021 June 9 between UT 7:50 and UT 12:00 in the \textit{R} band using the 1-meter Nickel telescope at Lick Observatory. The telescope is equipped with a $2\mathrm{K} \times 2\mathrm{K}$ CCD camera with a pixel scale of $0.184 \arcsec\, \rm{pixel}^{-1}$, resulting in a field of view of $ 6.0\arcmin\times6.0\arcmin$. 

The images were reduced and data was extracted as described in Section~\ref{Hall42}, with $2$ comparison stars and a photometric aperture of $20$ pixels ($3.68\arcsec$). The resulting light curve is presented in the middle right panel of Figure~\ref{fig:Transit0608}. The telescope was closed several times during the transit due to high humidity and rain, leading to data gaps in the light curve from 04:35:00 UT to 07:37:00 UT, from 08:40:00 UT to 09:36:00 UT, and from 9:58:00 UT to 10:27:00 UT. Although we successfully detected the transit, we did not include the resulting light curve into the global fitting in Section 4 because of the limited precision of the interrupted light curve.

\subsection{Photometry with Unistellar eVscopes}

We simultaneously observed the transit of WASP-148b with seven $4.5$-inch Unistellar eVscopes \citep{Marchis2020} operated by three professional astronomers (one of whom operated two eVscopes) and three amateur astronomers. Six were located in suburban and urban sites across California, and one was located in North Carolina. All eVscopes shared the same design: a Newtonian reflector with a 4.5-inch aperture and a Sony IMX224LQR CMOS sensor at the prime focus. The sensor has a pixel of scale 1.71\arcsec pixel$^{-1}$ and a $36.98'\times27.68'$ field of view, with a Bayer color filter array producing a mosaic of pixels with spectral responses peaking at blue, green, and red visible wavelengths. Individual images were recorded with exposure times of $3.97\,$s and digital sensor gains of 25 dB (0.129 e- ADU$^{-1}$).

Using the Unistellar/SETI Institute data reduction pipeline (SPOC) to process each data set individually, raw images were dark-subtracted and mutually aligned. These individual calibrated frames were then background-subtracted and averaged in groups of 30 into stacked images with 119.0 s of integration time each, increasing stellar S/N and time averaging over the pixel-dependent Bayer matrix response. Barycentric corrections to $\mathrm{BJD_{TDB}}$ were made to all image timestamps. Adaptively scaled elliptical apertures were then used to extract the fluxes of the target star and one comparison star to produce the differential light curve (via SExtractor's FLUX\_AUTO feature; \citealt{Bertin1996}).

Of the seven light curves obtained using eVscopes, we ultimately combined the three most complete light curves to provide our final light curve, shown in the rightmost panel of Figure \ref{fig:Transit0608}. Three other light curves were relatively incomplete due to imprecise pointing precision and/or cloud coverage, while the remaining telescope experienced technical issues. Before combination, the light curves were detrended for differential airmass extinction. 
Although we successfully detected the transit, we did not include the resulting light curve into the global fitting in Section 4 because of its limited precision.

\section{Stellar Parameters}
\label{section:stellar_parameters}

\subsection{Result from Keck/HIRES}
\label{section:stellar_parameters_keck}

We used the machine learning model \textit{The Cannon} \citep{ness2015cannon} following the procedure described in \citet{Rice2021} to extract stellar parameters from our iodine-free Keck/HIRES spectrum. Our training/validation set consists of the 1,202 FGK stars vetted in \citet{rice2020stellar}, drawn from the Spectral Properties of Cool Stars (SPOCS) catalog \citep{valenti2005spectroscopic}. This catalog includes 18 precisely determined stellar labels: 3 global stellar parameters (\teff, $\log g$, $v\sin i_{\star}$), and 15 elemental abundances: C, N, O, Na, Mg, Al, Si, Ca, Ti, V, Cr, Mn, Fe, Ni, and Y.

We determined the continuum baseline of each spectrum using the iterative polynomial fitting procedure outlined in \citet{valenti2005spectroscopic}. The continuum of each spectrum was divided out to produce a uniform set of reduced spectra each with a baseline flux at unity. These spectra and their associated normalized flux uncertainties were then input to \textit{The Cannon} for training and validation. We used the scatter of the validation set to determine the uncertainties inherent to the final trained model. Finally, we applied the trained model to the WASP-148 template spectrum in order to extract the stellar labels provided in Table \ref{table:results}. We report only the global stellar parameters and metallicity for direct comparison with the results obtained in Section \ref{section:stellar_parameters_EXOFASTv2}.

\subsection{Result from SED Fitting}
\label{section:stellar_parameters_EXOFASTv2}

We also used EXOFASTv2 \citep{Eastman2019} to estimate the stellar parameters of WASP-148, employing MESA Isochrones $\&$ Stellar Tracks (MIST) stellar
evolutionary models to perform a spectral energy distribution (SED) fit.

In this analysis, Gaussian priors were adopted for \teff, \feh, and stellar parallax. \teff\, and \feh\, were initialized using values derived in Section~\ref{section:stellar_parameters_keck}, and initial parallax values were drawn from corrected Gaia DR2 results \citep{Stassun2018}. To further constrain the host star radius, we enforced a upper limit on the $V$-band extinction from the Galactic dust maps \citep{Schlafly2011}.

With the MIST model, we fitted the SED of WASP-148 using broadband photometry provided by a series of photometric catalogs including Tycho-2, 2MASS All-Sky, AllWISE, and Gaia DR2. The resulting stellar parameters are shown in Table~\ref{table:results}. We note that the derived uncertainties in stellar physical parameters ($R_*$ and $M_*$) do not account for systematic errors in the data and are therefore underestimated \citep{Tayar2020}. The stellar atmospheric parameters (\teff, \feh, \logg, and $v\sin i_{\star}$) derived from Keck/HIRES and from the SED fit are in full agreement.




\end{document}